\documentclass[preprint,preprintnumbers,amsmath,amssymb,groupedaddress]{revtex4-1}
\usepackage{graphicx}
\usepackage{amssymb}
\usepackage{txfonts}
\begin{document}
%\preprint{}
\title{Structural Phase Transition, Antiferromagnetism and Two Superconducting Domes in LaFeAsO$_{1-x}$F$_{x}$ (0 $< x \leq$ 0.75)}
\author{J. Yang$^1$}
\author{T. Oka$^2$}
%\author{S. Kawasaki$^2$}
\author{Z. Li$^1$}
%\author{L. L. Wei$^1$}
\author{H. X. Yang$^1$}
\author{J. Q. Li$^1$}
%\author{Z. X. Zhao$^1$}
\author{G. F. Chen$^1$}
%\author{N. L. Wang$^1$}
\author{Guo-qing Zheng$^{1,2}$}
\affiliation{$^1$Institute of Physics, Chinese Academy of Sciences and Beijing National Laboratory for Condensed Matter Physics, Beijing 100190, China}
\affiliation{$^2$Department of Physics, Okayama University, Okayama 700-8530, Japan}

%\author{}
%\email[]{}
%\thanks{}
%\affiliation{}
\date{\today}

\begin{abstract}

We report $^{75}$As nuclear magnetic resonance (NMR) / nuclear quadrupole resonance (NQR) and transmission electron microscopy (TEM) studies on LaFeAsO$_{1-x}$F$_{x}$. There are two superconducting domes in this material. The first one appears at 0.03 $\leq$ $x$ $\leq$ 0.2 with $T_{\rm c}$$^{max}$ = 27 K, and the second one at 0.25 $\leq$ $x$ $\leq$ 0.75 with $T_{\rm c}$$^{max}$ = 30 K.
By NMR and TEM, we demonstrate that a $C4$-to-$C2$ structural phase transition (SPT) takes place above both domes, with the transition temperature $T_{\rm s}$ varying strongly with $x$. %, where fourfold rotation symmetry of the lattice is spontaneously broken.
In the first dome, the SPT is  followed by an antiferromagnetic (AF) transition, but neither AF order nor low-energy spin fluctuations are found in the second dome.
In LaFeAsO$_{0.97}$F$_{0.03}$, we find that AF order and superconductivity coexist microscopically via $^{75}$As nuclear spin-lattice relaxation rate (1/$T_1$) measurements.
%and suggest that the coexistence of antiferromagnetism and superconductivity is an universal feature for iron-based high $T_{\rm c}$ superconductors.
%The electronic phase diagram of the low doping region ($x \leq 0.15$) obtained by present measurements is different from previous report and similar to that of 122 system.
%Although the superconducting gap structure for 0.04 $\leq$ $x$ $\leq$ 0.15 is systematically understood by a sign-changed $s$-wave gap model with impurity scatterings,
%However, for $x$ = 0.03 where superconductivity and AF order coexists, and it is difficult to explain just by the framework of impurity scatterings.
In the coexisting region, 1/$T_1$ decreases  at $T_{\rm c}$ but becomes to be proportional to $T$ below 0.6$T_{\rm c}$, indicating
gapless excitations. Therefore, in contrast to the early reports, the obtained phase diagram for $x \leq$ 0.2 is quite similar to the doped BaFe$_{2}$As$_{2}$ system.
The electrical resistivity in the second dome can be fitted by $\rho = {{\rho }_{0}}+A{{T}^{n}}$ with $n$ = 1 and a maximal coefficient $A$ at around $x_{opt}$ = 0.5$\sim$0.55 where $T_{\rm s}$ extrapolates to zero and $T_{\rm c}$ is the maximal, which suggest the importance of quantum critical fluctuations associated with the SPT. We have constructed a complete phase diagram of LaFeAsO$_{1-x}$F$_{x}$, which provides  insight into  the relationship between  SPT, antiferromagnetism and superconductivity.

%These may provide another route to understand the mechanism of high-$T_{\rm c}$ superconductivity in iron-pnictide.
%$T_{\rm s}$ is suppressed like a second-order transition, which is much different from past result of La-1111.
%We have performed a systematic $^{75}$AsNuclear Quadrupole Resonance (NQR) measurements on iron-based superconductor LaFeAsO$_{1-x}$F$_{x}$.
%An antiferromagnetic(AF) order with $T_{\rm N}$ = 58K is observed and bulk superconductivity arises at $T_{\rm c}$ = 9.5K for $x$ = 0.03, which means the coexistence of AF order and superconductivity.
%The antiferromagnetic spin fluctuation (AFSF) found above $T_{\rm N}$ for $x$ = 0.03 persists in the regime 0.04 $\leq$ $x$ $\leq$ 0.08.
%AFSF is the strongest in $x$=0.04 and weakens with increasing $x$. For $x$ $\geq$ 0.10, AFSF is no longer observed.
%On the contrary, the superconducting transition temperature $T_{\rm c}$ shows a dome shaped $x$-dependence.
%These remind cuprates La$_{2-x}$Sr$_{x}$CuO$_4$, which suggests that the AFSF is important to produce the superconductivity in LaFeAsO$_{1-x}$F$_{x}$.
%In $x$ = 0.06, the spin-lattice relaxation rate ($1/T_1$) below $T_{\rm c}$ decreases exponentially down to 0.13 $T_{\rm c}$, which is clear evidence for a fully-gapped superconducting state.
%For $x$ either smaller or larger than 0.06, $1/T_1$ below $T_{\rm c}$ shows a nonexponential temperature dependence.
%Though they can be reproduced by the s$^{+-}$-wave model with impurity effect, much weaker $T$ dependence in $x$=0.03 is due to the coexistence of AF order and superconductivity.

\end{abstract}

%\pacs{74.25.Dw, 74.25.Nf, 74.70.-b}
%\keywords{}
\maketitle
%\section{}\label{}

\section{INTRODUCTION}%%%%%%%%%%%%%%%%%%%%%%%%%%%%%%%%%%%%%%%%%%%%%%%%%%%%%%%%%%%%%%%%%%%%%%%%%%%%%%%%%%%%%%%%%%%%%%%%%%%%%%%%%%%%%%%%%%%%%%%%%%%%

Iron-based superconductors (FeSCs) are a new class of high transition-temperature ($T_{\rm c}$) family\cite{Stewart}, which have attracted great interests in recent years. Vast efforts have been devoted to explore materials with higher $T_{\rm c}$\cite{ZARen,FeSefilm}, and to understand the  unconventional superconducting state\cite{Boeri,Mazin,Kuroki}. Soon after the first breakthrough of the discovery of the so-called "1111" structure LaFeAsO$_{1-x}$F$_{x}$\cite{Kamihara}, a wide variety of iron pnictides and chalcogenides, such as $A$Fe$_{2}$As$_{2}$ ("122")\cite{Rotter}, $A$FeAs ("111")\cite{111_jin}, FeSe ("11")\cite{Hsu} were discovered successively.
Most FeSCs show a tetragonal-orthorhombic structural phase transition (SPT) at $T_{\rm s}$ followed by an antiferromagnetic (AF) order at $T_{\rm N}$. Element substitution or external pressure suppress both $T_{\rm s}$ and  $T_{\rm N}$ then lead to superconductivity\cite{Stewart}.
Since unconventional superconductivity emerges in close proximity to antiferromagnetism, the AF spin fluctuations are naturally proposed to be responsible for the electron pairing\cite{Davis}, as is the situation in heavy fermions\cite{Lonzarich} and high-$T_{\rm c}$ cuprates\cite{cuprates}.
More recently, the electronic nematicity, a phenomenon of spontaneous rotation-symmetry breaking in the Fe-plane below $T_{\rm s}$, has emerged as another hot research topic in FeSCs  \cite{Fernandes-NatPhys}.
Electrical resistivity\cite{JHChu}, spin excitation\cite{Lu,Zhou-PRB} and magnetic torque\cite{Kasahara}  show large in-plane anisotropy. Such electronic nematicity may stem from the band splittings of the Fe-3$d_{xz}$ and 3$d_{yz}$ orbitals\cite{Yi}.
%It is believed by many that  the electronic nematicity is the  driving force of the structural phase transition, and may play an important role
%in  both antiferromagnetism and superconductivity\cite{Fernandes-NatPhys}.
Many intriguing properties arising from  nematic fluctuations associated with a  quantum critical point (QCP)  has been reported\cite{Zhou,Fisher-Science}. However, the origin of  nematic order is still controversial;  both spin-\cite{Fernandes-PRL} and orbital-\cite{Onari,Chen,Lee,Philip}  scenario have been proposed.
Thus, antiferromagnetism and electronic nematicity below $T_{\rm s}$ are two noteworthy characteristics of FeSCs, which hold clues to the underlying of the physics in this new class of materials.
%
%The phase diagram of FeSCs exhibits different orders that  are closely related to superconductivity, therefore, one or more quantum critical points (QCPs) could be accessed by tuning parameters.
%Their relation with unconventional superconductivity remains the most attractive issue.

After the AF order is suppressed,  spin fluctuations have been reported for different 122 systems\cite{Ning,Li2,nakai0,Zhou}, and also in LaFeAsO$_{1-x}$F$_{x}$ with $x \leq 0.15$\cite{Oka}. However, so far the phase diagram for the prototypical FeSC LaFeAsO$_{1-x}$F$_{x}$ in the underdoped region is still unclear.
Early works suggested that $T_{\rm s}$ and $T_{\rm N}$ stay constant with increasing doping and %of LaFeAsO$_{1-x}$F$_{x}$
vanish abruptly at some doping level, before superconductivity emerges \cite{Luetkens,Huang}. Also, %On the other hand,
spatial phase separation of AF and paramagnetic-superconducting domains were reported near the phase boundary %of LaFeAsO$_{1-x}$F$_{x}$
\cite{Fujiwara,hiraishi}.
Such phase diagram is quite different from that of other FeSCs, for example, CeFeAsO$_{1-x}$F$_{x}$\cite{zhao}, SmFeAsO$_{1-x}$F$_{x}$\cite{Drew2} and 122 systems\cite{Hashimoto,Zhou,Li2,nakai0}, where $T_{\rm s}$ and $T_{\rm N}$ decrease  with increasing doping. Furthermore, superconductivity coexists microscopically with AF state in 122 systems\cite{Zhou,Li1}.
It is unclear whether the early-reported properties are due to the poor polycrystalline sample quality  or arise from the intrinsic property of LaFeAsO$_{1-x}$F$_{x}$.

As for the carrier doping, early studies suggested that the F-content can not exceed 0.2\cite{Kamihara,Luetkens,Oka}. However, by high-pressure synthesis technique, we recently found that the F-content can go as high as 0.75\cite{yang}. In the heavily doped region, we discovered another superconductivity dome centered at $x$ = 0.55 with an even higher $T_{\rm c}$ = 30 K\cite{yang}. Surprisingly, we found that a structural phase transition takes place above the new dome\cite{yang}. This discovery raised interest on the connection to the SPT in the low doping region, and on the role of the electronic state change below $T_{\rm s}$ in FeSCs in a broader context.

In this paper, we address the issues of the SPT in the two doping regimes. We also attempt to construct a complete phase diagram of the low doping region, and explore the interplay between antiferromagnetism and superconductivity. By $^{75}$As NMR and TEM, we demonstrate that the structural phase transitions taking place in the low-doped and high-doped regimes share similarities. %characteristic.
%By $^{75}$As NMR, we identify the structural phase transition at $T_{\rm s}$ = 135 K, 100 K and 250 K for $x$ = 0.03, 0.04, 0.55, respectively. Meanwhile, we have directly confirmed $T_{\rm s}$ by TEM.
In the low doping region, the suppression of $T_{\rm s}$ and $T_{\rm N}$ shows a second-order-like variation towards the first superconducting dome.
For $x$ = 0.03, a long range AF order at $T_{\rm N}$ = 58 K with a magnetic moment of $m_{\rm Fe}$ $\sim$ 0.011 $\mu_{\rm B}$ is found and bulk superconductivity sets in at $T_{\rm c}$ = 9.5K. The measurement of the spin-lattice relaxation rate (1/$T_1$) indicates a microscopic coexistence of AF order and superconductivity.
Our results show that the phase diagram of LaFeAsO$_{1-x}$F$_{x}$ in the low-doped regime ($x \leq 0.15$) is similar to that of the 122 systems.
%Furthermore, the $^{75}$As nuclear spin-lattice relaxation rate 1/$T_1$ for $x$ = 0.03 decreases in proportional to almost $T$ below $T_{\rm c}$, which indicates that there exists a large residual density of states (RDOS) in superconducting gaps.
%%Although the temperature dependence of 1/$T_1$ for 0.04 $\leq$ $x$ $\leq$ 0.15 can be understood by s$^{\pm}$-wave model with impurity scattering, it is hard to attribute weak $T$ dependence of 1/$T_1$ for $x$ = 0.03 to impurity scattering.
%Our results indicate the presence of a
%magnetic quantum critical point.
In the second dome, however, neither AF order nor spin fluctuation can be found, but $T_{\rm s}$ increases with increasing $x$ for $x$ $>$ 0.5. $T_{\rm s}$ extrapolates to zero at around $x_{opt}$ = 0.5$\sim$0.55, where the electrical resistivity shows a $T$-linear behavior and the coefficient $A$ from the $\rho = {{\rho }_{0}}+A{{T}^{n}}$ fitting shows a maximum. These interesting properties may originate from the quantum fluctuation associated with a nematic order.

This paper is organized as following. The experimental methods  are described in Sec. \ref{method}. In section \ref{AF}, the $^{75}$As NQR spectra that evidence an AF order are presented. Evidence for a coexistence of AF order and superconductivity are shown in Sec. \ref{Coexistence}. Section \ref{Structure} discusses the structural phase transitions in the two doping regimes, on the basis of TEM and NMR data. Finally, a possible new type of quantum criticality in the second dome is discussed in section \ref{QCP}.

\section{EXPERIMENTAL METHODS}\label{method}

The polycrystalline LaFeAsO$_{1-x}$F$_{x}$ samples were prepared by the two-step solid state reaction method. Here, $x$ indicates the nominal composition of the starting material. In the first step, the precursor LaAs powder was obtained by reacting La pieces (99.5\%) and As powders (99.999\%) at 500$^{\circ}$C for 12 hours then at 850$^{\circ}$C for 2 hours. In the second step, samples with different fluorine concentrations were sintered under ambient pressure (AP) and high pressure (HP), respectively.
We adopted the AP method for samples with $x$ = 0.03 - 0.2~\cite{GFChen}. The stoichiometric mixtures of the starting materials LaAs, Fe$_{2}$O$_{3}$, Fe, and LaF$_{3}$ were ground thoroughly and cold-pressed into pellets.
The pellets were placed into Ta crucible and sealed in quartz tube.
%All these  processes were performed in a grove box filled with high purity argon gas and the concentration of oxygen and/or water was kept less than 1 ppm.
They were then sintered at a temperature of 1150$^{\circ}$C for 50 hours.
%Different from the synthesis method reported by Kamihara $et$ $al$\cite{Kamihara},
%Fe$_{2}$O$_{3}$ is used as a source of oxygen.%instead of La$_{2}$O$_{3}$.
LaFeAsO$_{1-x}$F$_{x}$ with $x$ = 0.25 - 0.75 were synthesized by the HP method. The starting materials LaAs, Fe, Fe$_{2}$O$_{3}$ and FeF$_{2}$ were mixed together according to the nominal ratio and pressed into pellets. Different from the AP method, the pellets were sealed in boron nitride crucibles and sintered in a six-anvil high-pressure synthesis apparatus under a pressure of 6 GPa at 1250$^{\circ}$C for 2-4 hours. After sintering, the sample was quenched to room temperature by water cooling within a few seconds, and then the pressure was released.
%This rapid quenching process was not used in the previous works \cite{Ren,Lu}, but is important %was supposed to keep the meta-stable phase which cannot be  formed at ambient pressure.

Compared to the solid state reaction under ambient pressure, the high pressure synthesis method have two advantages. Fist, the raw materials are sealed  and pressurized in the whole synthesis process, so fluorine element, which is  volatile and easily react with silica, can be kept. Second, reaction under high pressure and the rapid quenching process help to keep the meta-stable phase that can not be  formed at ambient pressure.
%All above  procedures  treating raw materials were done in an Ar-protected glove box.

Powder x-ray diffraction (XRD) with Cu K$\alpha$ radiation ($\lambda$= 0.154nm) were performed at room temperature to characterize the phase purity and structural parameters. The temperature dependence of resistivity were measured by a standard four-probe method. %The electrical terminals were made by silver paste.
The value of $T_{\rm c}$  was determined
by both dc susceptibility using a superconducting
quantum interference device (Quantum Design)
and ac susceptibility using an $in$-$situ$ coil.
%We define $T_{\rm c}$ of AP samples by the ac susceptibility measurements using the $in-situ$ coil \cite{Oka}, which indicates $T_{\rm c}$ = 9.5, 21, 27, 23, 18, and 12 K for $x$ = 0.03, 0.04, 0.06, 0.08, 0.10, and 0.15, respectively.
%For HP samples, DC susceptibility  determines $T_{\rm c}$ to be 17.5, 20.5, 26, 29.5, 30, 28.5, 27, 24 K for $x$ = 0.25, 0.3, 0.4, 0.5, 0.55, 0.6, 0.65 and 0.75, respectively.
%For NMR/NQR measurements, the obtained polycrystalline samples were crushed into powders.
$^{75}$As NMR/NQR measurements were carried out by using a phase-coherent spectrometer. The NMR spectra were obtained by scanning the frequency and integrating the spin echo at a fixed magnetic field $H_{0}$. The NQR spectra were also taken by changing the frequency point by point.
The spin-lattice relaxation time $T_{1}$ was measured by using the saturation-recovery method.
The recovery curve of $^{75}$As ($I$ = 3/2) NQR is well fitted by a single exponential function  $1-M(t)/M_{0} = \exp(-3t/T_{1})$, %while the recovery curve of $^{75}$AsNMR follows $1-M(t)/M(\infty) = 0.1exp(-t/T_{1}) + 0.9exp(-6t/T_{1})$ for the central transition peak,
where $M_0$ and $M(t)$ are the nuclear magnetization in the thermal equilibrium and at a time $t$ after the saturating pulse, respectively \cite{Narath}.

Specimens for TEM were prepared by crushing the bulk material into fine fragments which were then supported by a copper grid coated with a thin carbon film. A JEOL 2100F TEM, equipped with cooling (below $T$ = 300 K) or heating sample holders (above $T$ = 300 K), was used for investigating the structural properties of the samples.

\section{Results}

\subsection{Magnetic order in the low doping region}\label{AF}%%%%%%%%%%%

\begin{figure}[htbp]
\includegraphics[width=10cm]{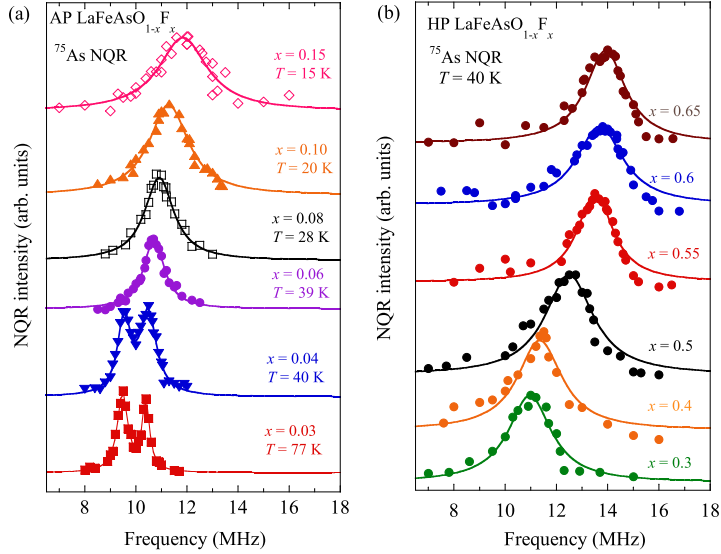}% Here is how to import EPS art
\caption{(color online)
(a)(b) $^{75}$As NQR spectra for AP LaFeAsO$_{1-x}$F$_{x}$ with 0.03 $\leq x \leq$ 0.15 and HP samples with 0.3 $\leq x \leq$ 0.65, respectively.
The solid curves are Lorentzian function fittings.
\label{AsNQR_all}
}
\end{figure}

%In the low doping region of  LaFeAsO$_{1-x}$F$_{x}$, superconductivity emerges adjacent to an AF order,
%Their superconductivity coexists microscopically with AF state and it is thus considered that the spin fluctuations associated with magnetic  Quantum Critical Point (QCP) are crucial for the mechanism of superconductivity.
%We have reported that AF spin fluctuation is also important to induce the superconductivity in LaFeAsO$_{1-x}$F$_{x}$\cite{Oka}.
%
%In the low doping region of LaFeAsO$_{1-x}$F$_{x}$, superconductivity emerges adjacent to an AF order, thus the evolution of the AF order with F-content $x$ is a crucial topic for underdtanding superconductivity.

We first present the results for  AF order  in the low doping region revealed by $^{75}$As NQR.
%Previously, we have reported the $^{75}$As NQR spectra for 0.03 $\leq x \leq$ 0.15 measured above $T_{\rm c}$\cite{Oka}.
%In this work, we also perfromed $^{75}$As NQR measurements for HP  samples with 0.3$\leq x \leq$0.65  at $T$ = 40 K.
The $^{75}$As NQR spectra for AP and HP samples are shown in Fig. \ref{AsNQR_all}(a),(b), respectively.
A clear single peak, which can be fitted by a single Lorentzian function, was observed for $x \geq$ 0.06.
Theoretically, $^{75}$As NQR has only one peak corresponding to the $m = \pm1/2 \leftrightarrow \pm3/2$ transition, and  the NQR frequency $\nu_Q$ probes the  electric field gradient (EFG) generated by the carrier distribution and the lattice contribution surrounding the target nucleus.
Thus the well-resolved NQR spectra indicate that the carrier doping distribution and the lattice surroundings at As site are uniform for $x \geq$ 0.06.
In contrast, two peaks were observed for $x$ = 0.03 and 0.04, which means that there exist two As sites with different EFG surroundings.
This may due to the local arrangement of the F ion in the underdoped samples.
Similar $^{75}$As NQR spectra of two peaks in underdoped LaFeAsO$_{1-x}$F$_{x}$ were also reported by other groups\cite{Lang}.
In what follow, we denote the lower (higher) frequency peak with "Low" ("High").

\begin{figure}[htbp]
\includegraphics[width=6cm]{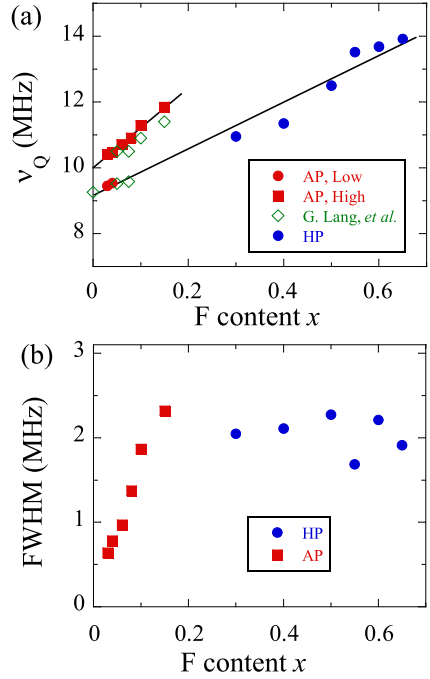}% Here is how to import EPS art
\caption{(color online)
(a)(b) The $x$ dependence of the $^{75}$As NQR frequency $\nu_Q$ and the full width at half maximum (FWHM) of the spectra, respectively. The solid lines are guide to the eyes.
\label{AsNQRFWHM}
}
\end{figure}

The obtained doping dependence of the $^{75}$As NQR frequency $\nu_Q$ is shown in Fig. \ref{AsNQRFWHM} (a). $\nu_Q$ increases almost linearly with increasing the nominal $x$ content. This result together with the fact  that the lattice constant obtained by XRD also changes continuously as $x$ increases\cite{yang2} ensure that the carrier content does increase with increasing $x$.
The full width at half maximum (FWHM) of the $^{75}$As NQR spectra are shown in Fig. \ref{AsNQRFWHM} (b).
Since the distribution of the F-content will result in a broadening of the NQR spectrum, it is reasonable that the FWHM increases with increasing $x$ for AP samples. The FWHM of the HP samples is almost $x$ independent and  comparable to that of $x$ = 0.1 grown at ambient pressure, which indicates that high-pressure synthesis does not bring about additional F-content distribution.

\begin{figure}[htbp]
\includegraphics[width=8cm]{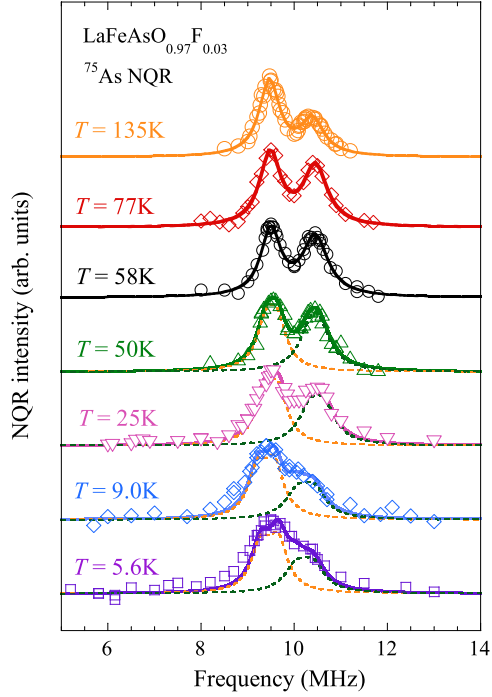}% Here is how to import EPS art
\caption{\label{AsNQRspectum_x=0.03} (color online)
(a) The temperature dependence of the $^{75}$As NQR spectra for LaFeAsO$_{0.97}$F$_{0.03}$.
Solid curves above 58K are fittings to two Lorentzians. Below 58K, solid and dotted curves are simulations as described in the text.
}
\end{figure}

\begin{figure}[tbp]
\includegraphics[width=7.5cm]{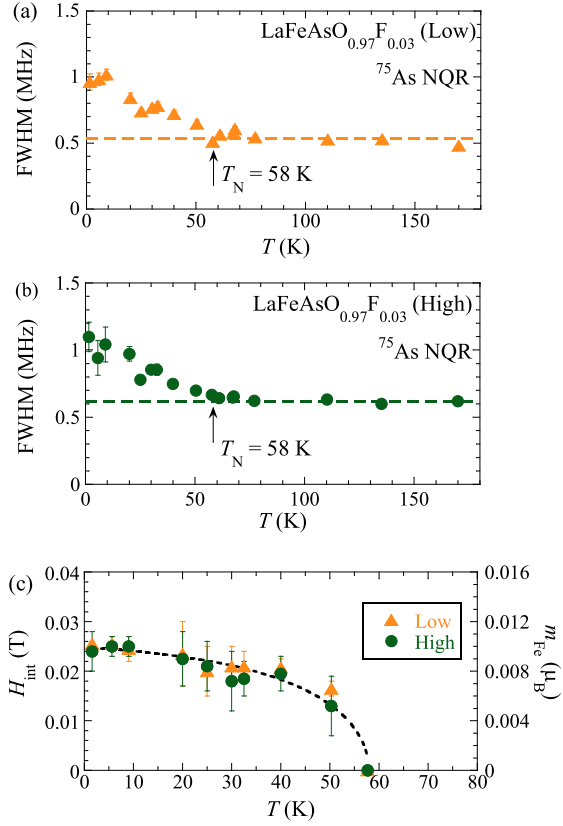}
\caption{(color online)
(a)(b) The temperature dependence of FWHM for the two NQR peaks.% at lower and  higher frequency, respectively.
(c) The temperature dependence of the internal magnetic field ($H_{int}$) at As site (left vertical axis) and the estimated Fe moment for Low and High (right vertical axis). Dashed lines and  curves are guides to the eyes.
%(c) The temperature dependence of internal magnetic field at As-site obtained by the simulation.
%The solid curve indicates the molecular-field theory with $s$=1/2 and $H_{int}$ $\propto$ $m$=tanh($mT_N/T$).
\label{AsNQRFWHM_x=0.03}
}
\end{figure}

Figure \ref{AsNQRspectum_x=0.03} shows the temperature dependence of the $^{75}$As NQR spectra for $x$ = 0.03.
It is obvious that the spectra are broadened at low temperatures.
Figure \ref{AsNQRFWHM_x=0.03}(a) and (b) show the temperature dependence of the FWHM for Low and High peaks obtained by a two-Lorentian fitting.
For both of Low and High peaks, FWHM increases below 58 K.
In the following, we elaborate that the broadening of NQR spectrum is due to an AF order.
The nuclear spin Hamiltonian which derives from the nuclear quadrupole interaction is given by\cite{Abragam}
\begin{equation}
\mathcal{H}_{Q}=\frac{eQV_{zz}}{4I(2I-1)}((3\hat{I}_{z}^{2}-\hat{I}^{2})+\eta (\hat{I}_{x}^{2}+\hat{I}_{y}^{2})),
\label{HQ}
\end{equation}
where $eQ$ is the electric quadrupole moment, $V_{\alpha\beta}$ is the EFG tensor, and $\eta=|V_{xx}-V_{yy}|/V_{zz}$ is the asymmetry parameter of the EFG.
For $^{75}$As nucleus ( $I$ = 3/2 ), the $\pm$1/2 $\leftrightarrow$ $\pm$2/3 transition gives rise to a peak at $\nu_Q=\frac{eQV_{zz}}{2h}\sqrt{\mathstrut 1+\eta^{2}/3}$.
When an AF order occurs and an internal magnetic field sets in, the Hamiltonian will be perturbed by the Zeeman interaction. The perturbative Hamiltonian is given by $\mathcal{H}_{Z}=-\gamma \hbar \vec{I}\cdot\vec{H}_{int}$, where $\gamma$ is the gyromagnetic ratio and $\vec{H}_{int}$ is the internal magnetic field, respectively.
Since the direction of $\vec{H}_{int}$ at As-site is parallel to the $c$ axis in LaFeAsO$_{1-x}$F$_{x}$\cite{Kitagawa}, the perturbation can be written as $\mathcal{H}_{Z}=-\gamma\hbar\hat{I}_{z}H_{int}$.
This perturbation removes the degeneracy of energy levels and the single $^{75}$As NQR peak will split into three peaks, corresponding to -3/2 $\leftrightarrow$ -1/2, -1/2 $\leftrightarrow$ 1/2, and 1/2 $\leftrightarrow$ 3/2 transitions. In particular, -3/2 $\leftrightarrow$ -1/2 and 1/2 $\leftrightarrow$ 3/2 transitions locate at $\nu_Q+\frac{\gamma}{2\pi}H_{int}$ and $\nu_Q-\frac{\gamma}{2\pi}H_{int}$, respectively.
For our case in $x$ = 0.03,  the NQR spectra do not split completely due to a small $H_{int}$; the two transitions overlap, resulting in a broad peak. As shown shown in Fig. \ref{AsNQRFWHM} (b), the FWHM increases rapidly below $T_{\rm N}$.

To estimate the value of the internal magnetic field at As site, we performed a simple simulation.
Since FWHM of both Low and High increases below $T_{\rm N}$ = 58 K, we assumed that different size of the internal magnetic field is produced at Low and High sites. For each As site, we have reproduced the spectra using two Lorentzians.
%We assumed that FWHM of Lorentians for Low and High are 0.549MHz and 0.615MHz, respectively, which were averaged values above $T_{\rm N}$.
Figure \ref{AsNQRspectum_x=0.03} shows the spectra below $T_{\rm N}$ with the simulations.
Figure \ref{AsNQRFWHM_x=0.03} (c) shows the temperature dependence of $H_{int}^{Low}$ and $H_{int}^{High}$ obtained by this simulation.
Since the internal fields of Low and High have nearly the same values and temperature-variation trend, the AF order occurs homogeneously in the $x$ = 0.03 sample. %, with identical $T_{\rm N}$ in Low and High.
  %the electronic state is homogeneous,
%Although the NQR spectrum for $x$ = 0.03 shows
The appearance of two peaks %,  and $T_{\rm c}$ are identical ,
 may be understood as NQR being sensitive to the local dopant arrangement. Similar results of multiple NQR peaks with identical $T_{\rm N}$ were also observed in heavy-fermion CeRh$_{1-x}$Ir$_{x}$In$_{5}$ \cite{Zheng}.

Furthermore, we have estimated the magnetic moment of ordered Fe atom by using $H_{int}$($^{75}$As)=$^{75}A_{hf}m_{\rm Fe}$, where the Hyperfine coupling constant $^{75}A_{hf}$=25 kOe/$\mu_{\rm B}$ is taken from other NMR measurement\cite{grafe1}.
The obtained magnetic moment saturates at low temperatures with  $m_{\rm Fe}$ $\sim$ 0.011 $\mu_{\rm B}$, which is much smaller than that of parent compound of 0.36$\mu_{\rm B}$ obtained from neutron scattering\cite{cruz}.
As increasing the  electron doping, the Fe magnetic moments are suppressed significantly.
The broadening of NQR spectrum is not observed for $x$ $\geq$ 0.04, which indicates that the AF order vanishes between $x$ = 0.03 and 0.04.

\subsection{Coexistence of the AF order and superconductivity}\label{Coexistence}%%%%%%%%%%%

\begin{figure}[htbp]
\includegraphics[width=8cm]{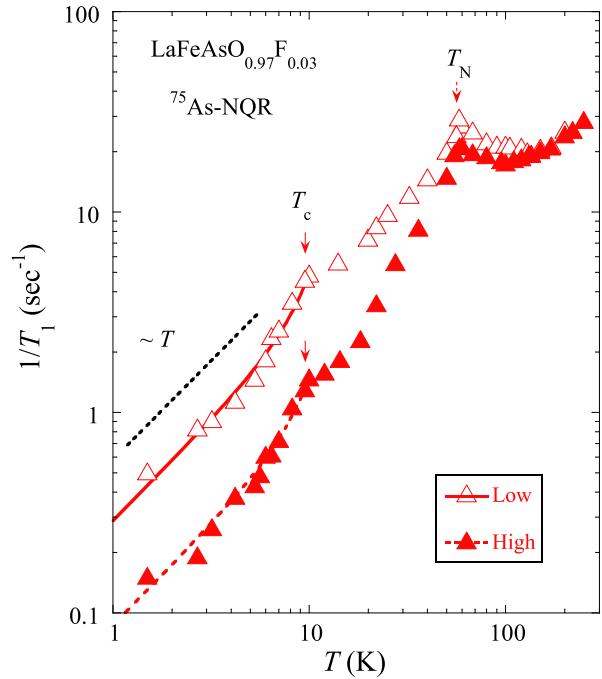}
\caption{(color online) The temperature dependence of the spin-lattice relaxation rate (1/$T_1$) for $x$ = 0.03 measured at High and Low peaks, respectively.
Solid and dashed curves below $T_{\rm c}$ are guides to the eyes.
Dotted line indicates the relation 1/$T_1$ $\propto$ $T$.
Dotted and solid arrows indicate $T_{\rm N}$ and $T_{\rm c}$, respectively.
\label{T1andRecovery}
}
\end{figure}

\begin{figure}[htbp]
\includegraphics[width=5cm]{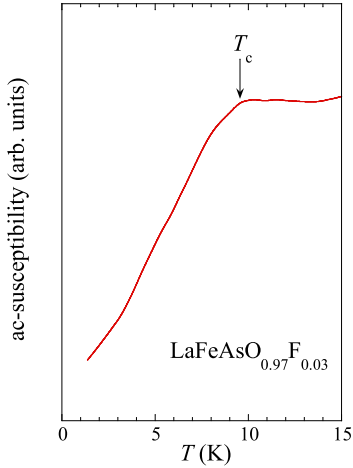}
\caption{(color online) The AC susceptibility for LaFeAsO$_{0.97}$F$_{0.03}$ measured by the $in$-$situ$ NQR coil.
\label{AC_Chi}
}
\end{figure}

The interrelation between antiferromagnetism and superconductivity is one of the most intriguing issues.
In this section, we present the experimental evidence for the microscopic coexistence of the AF order and superconductivity.
We measured the spin-lattice relaxation rate (1/$T_1$) at Low ($f$ = 9.5 MHz) and High ($f$ = 10.3 MHz) for $x$ = 0.03.
The nuclear magnetization recovery curves of both Low and High peaks are of single component. Figure \ref{T1andRecovery} shows the temperature dependence of $1/T_{1}$.
For both Low and High, $1/T_1$ forms a peak at $T_{\rm N}$ = 58 K due to a critical slowing down of the magnetic moment. As the temperature is reduced, $1/T_1$  decreases steeply at $T$ = 9.5 K. The AC susceptibility measured by $in$-$situ$ NQR coil shows that diamagnetism shows up below  this temperature ($T_{\rm c}$ = 9.5 K), thus  the sharp decrease in $1/T_1$ is due to the opening of a superconducting gap (Fig. \ref{AC_Chi}).
The results of NQR spectra and 1/$T_1$ indicate that superconductivity coexists microscopically with AF order in LaFeAsO$_{0.97}$F$_{0.03}$.

%In the previous paper, we suggested that

\begin{figure}[htbp]
\includegraphics[width=8cm]{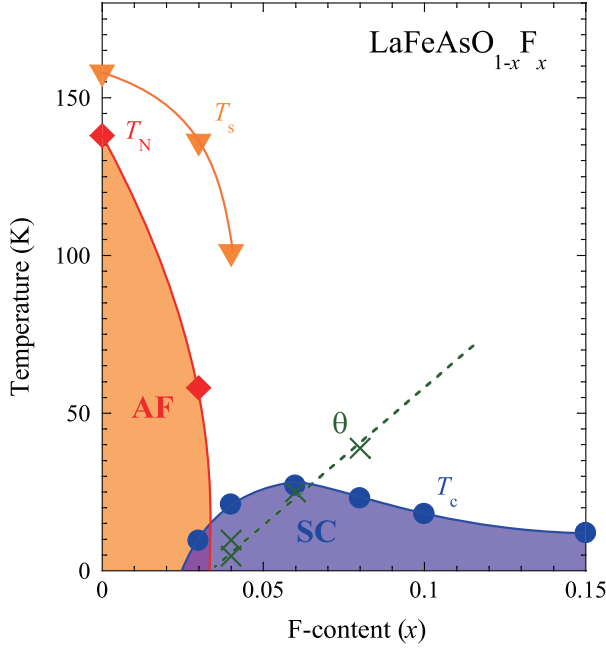}
\caption{(color online) The Phase diagram for LaFeAsO$_{1-x}$F$_{x}$ in the low-doped regime ($x$ $\leq$ 0.15). AF and SC denote the antiferromagnetic ordered state and superconducting state, respectively.
The deep purple area indicates the state where AF order and superconductivity coexist. $T_{\rm s}$ and $T_{\rm N}$ for $x$ = 0 are referred from ref.\cite{Huang}. For the estimate of $T_{\rm s}$, see Sec. \ref{Structure} of the main text. The Weiss temperature $\theta$ is obtained
from fitting the $1/T_1$ data to  the theory of weakly antiferromagnetically-correlated metal, $1/T_{1}T=(1/T_{1}T)_{0}+C/(T+\theta)$\cite{moriya1}. The dotted line is a guide to the eyes.
}
\label{phasediagram_c}
\end{figure}

Compared to the parent compound LaFeAsO, $T_{\rm N}$ for $x$ = 0.03 is suppressed greatly. Our results of NQR spectra suggest that the AF order vanishes between $x$ = 0.03 and 0.04. These features were not seen at all in the previous works\cite{Luetkens,Huang}, which report that $T_{\rm N}$ only decreases slightly by F doping and vanishes abruptly at some doping level.
For $x$ = 0.03 in our case, the small moment $m_{\rm Fe}$ = 0.011 $\mu_{\rm B}$ is probably a factor in favor of the coexistence of  antiferromagnetism and superconductivity.

The evolution of $1/T_{1}T$ with doping level $x$ suggests that, with doping, the system  is approaching a magnetic instability between $x$ = 0.03 and 0.04. % (see Sec. \ref{QCP} for $1/T_{1}T$ and the fittings).
According to the theory of weakly antiferromagnetically-correlated metal, $1/T_{1}T$ is proportional to the staggered magnetic susceptibility $\chi^{''}(q)$ and follows a Curie-Weiss law\cite{moriya1}, $1/T_{1}T=(1/T_{1}T)_{0}+C/(T+\theta)$.
Here, the first term is the contribution from the density of states at the Fermi level, and the second term describes the contribution from the antiferromagnetic wave vector $Q$. The $1/T_{1}T$ data can be well fitted by this theory \cite{Oka}.
The parameter $\theta$ approaches to 0 K between $x$ = 0.03 and 0.04,  %as shown in
which means that the  $\chi^{''}(Q)$ diverges at $T$ = 0 K there.
These facts suggest the existence of a  magnetic QCP between $x$ = 0.03 and 0.04.
The obtained phase diagram at low-doped regime is shown in Fig. \ref{phasediagram_c}. It shares many similarities with  the 122 system.

Next, we turn to the superconducting state in the coexistence region. Below $T_{\rm c}$, $1/T_{1}$ for $x$ = 0.03 decreases, but becomes almost proportional to $T$ below 5.7 K.
%, which is quite different from that of the samples away from the coexistence region.
This is in contrast to the  $x$ = 0.06 sample with the highest $T_{\rm c}$ \cite{Oka} or the optimally-doped Ba$_{1-x}$K$_{x}$Fe$_2$As$_2$ \cite{Li2},  where $1/T_{1}$ below $T_{\rm c}$ decreases exponentially. %, as was found in .
%This indicates multiple fully-opened superconducting gaps, and can be understood by s$^{\pm}$-wave symmetry model\cite{Mazin,Kuroki} with impurity scatterings.
%Although the temperature dependence of 1/$T_1$ for $x$=0.03 can be reproduced also by this model, the impurity scattering parameter $\eta$ is about 10 times larger than that of $x$ = 0.04\cite{Oka}.
The behavior below $T_{\rm c}$ seen in $x$ = 0.03 sample can not be ascribed to impurity scattering, since the line width of NQR spectrum for $x$ = 0.03 is smaller than that for $x$ = 0.06.
Similar $T$-linear behavior of 1/$T_1$ %below $T_{\rm c}$
was also observed in underdoped  Ba$_{0.77}$K$_{0.23}$Fe$_2$As$_2$\cite{Li1} and Ca$_{1-x}$La$_{x}$FeAs$_{2}$\cite{Kawasaki_112}, where superconductivity coexists with AF order.
The gapless state in the coexistence region deserves further study. One possibility is that it arises from  the excitations of  an exotic pairing state with mixed spin-triplet component due to   the coexisting magnetism \cite{Chubukov}.

\subsection{Structural phase transition}\label{Structure}%%%%%%%%%%%%%

In the parent compound LaFeAsO, a structural phase transition takes place  above $T_{\rm N}$\cite{cruz}, but the evolution of $T_{\rm s}$ with F content $x$ is unclear in the low-doped regime.
% for 0 $<$ $x$ $<$ 0.15.
In the high doping regime of $x$ $>$ 0.2, we found that a structural phase transition also occurs, with $T_{\rm s}$ intersecting  the new superconducting dome. In this section, we compare the structural phase transition in the two doping regimes. %for AP LaFeAsO$_{1-x}$F$_{x}$ ($x$ $<$ 0.15) with that for HP samples ($x$ $>$ 0.5), and explain how we determine $T_{\rm s}$.

\begin{figure}[htbp]
\includegraphics[width=8cm]{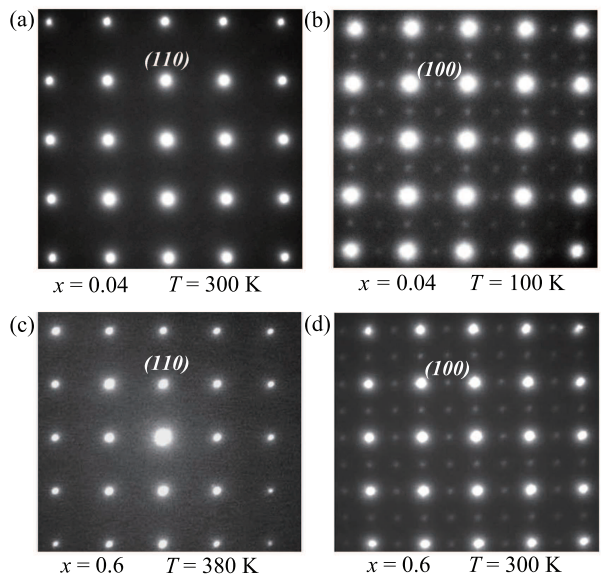}
\caption{
The TEM images for LaFeAsO$_{0.96}$F$_{0.04}$ and LaFeAsO$_{0.4}$F$_{0.6}$.
}
\label{TEM}
\end{figure}

First, we directly confirmed the structural phase transition   by TEM images. Figure \ref{TEM}(a)(b) show the  [001] zone-axis electron diffraction patterns for  LaFeAsO$_{0.96}$F$_{0.04}$ taken at $T$ = 300 K and $T$ = 100 K. At room temperature, only (110) spots can be seen, which indicates that the crystal structure is tetragonal with the space group of $P$4/$nmm$. At $T$ = 100 K, additional spots appear at (100) positions, which means the $C4$ crystal symmetry is lowered. These features  are similar with the TEM results of  the HP LaFeAsO$_{1-x}$F$_{x}$\cite{yang}. Figure \ref{TEM}(c)(d) present an example for $x$ = 0.6. At $T$ = 300 K the (100) spots already exist, indicating a broken $C4$ symmetry. Upon heating from room temperature, the (100) spots
disappeared at $T$ = 380 K. Therefore $T_{\rm s}$ = 380 K was identified for this composition. From the point of view of TEM, structural phase transition is similar between AP and HP samples.

\begin{figure*}[htbp]
\includegraphics[width=12.0cm]{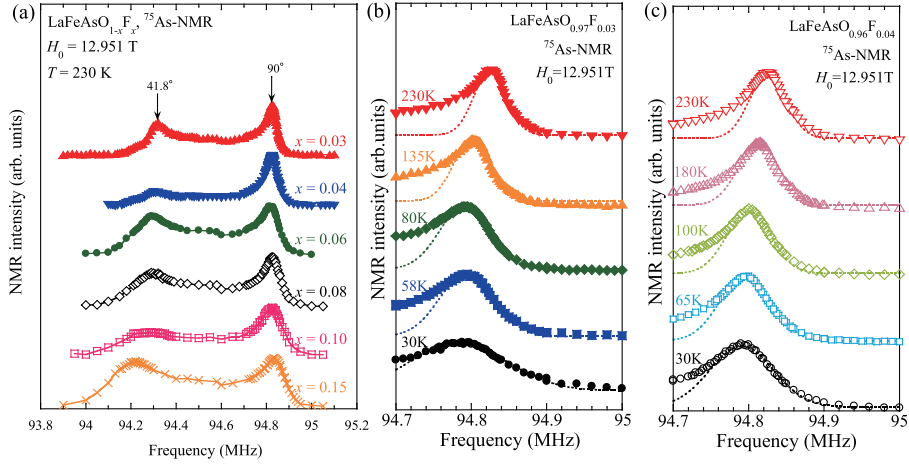}
\caption{(color online) (a) Doping dependence of the frequency-swept $^{75}$As NMR spectrum (center peak only) at the fixed magnetic field of $H_0$ = 12.951 T for LaFeAsO$_{1-x}$F$_{x}$ ($x$ = 0.03-0.15).
The two horns correspond to $\theta$ = 41.8$^\circ$ and 90$^\circ$, respectively.
(b)(c) Temperature dependence of the $^{75}$As NMR spectrum for $x$ = 0.03 and $x$ = 0.04.  The dotted lines are Gaussian-function fittings.
}
\label{AsNMRspectrum}
\end{figure*}

Next, we present the results for structural phase transition seen by $^{75}$As NMR.
Figure \ref{AsNMRspectrum} (a) shows the $^{75}$As NMR spectra for AP samples with 0.03 $\leq$ $x$ $\leq$ 0.15 measured at $T$ = 230K.
For $^{75}$As NMR,
%because $^{75}$As has a nuclear quadrupole moment, the nuclear quadrupole interaction with local EFG in the crystal acts as a perturbation.
the total nuclear spin Hamiltonian is given by \cite{Abragam}
\begin{equation}
\mathcal{H}= \gamma \hbar H_{0}(1+K)\hat{I}_{z'} +\frac{h\nu_{Q}}{6}[(3\hat{I}_{z'}^{2}-\hat{I}^{2})+\eta(\hat{I}_{x'}^{2}-\hat{I}_{y'}^{2})],
\label{Hz}
\end{equation}
where the first term is from the Zeeman interaction with $K$ being the Knight shift, and the second term represents the interaction of the nuclear quadrupole moment with EFG tensor.
In the high-field limit, the quadrupolar term can be treated as a perturbation. The principle axes ($x'$, $y'$, $z'$) of the EFG are determined by the local symmetry in the unit cell. $\theta$ is the angle between the applied field $H_{0}$ and the $z'$ axis.
%The resonance frequency of a particular nuclear transition depends on field direction relative to the crystalline axes.
In the case of random powder samples with a uniform distribution of $\theta$,
the central transition ($I_z$= -1/2 $\leftrightarrow$ 1/2) of $^{75}$As NMR will show a characteristic shape called "powder pattern". It can be seen from Fig. \ref{AsNMRspectrum} (a) that all samples show a two-horns shape, as expected for a powder pattern, where the lower frequency horn and higher frequency horn correspond to $\theta$ = 41.8$^\circ$ and 90$^\circ$, respectively.
For LaFeAsO$_{1-x}$F$_{x}$, the principle axes $x'$, $y'$, $z'$ of the EFG coincide with the crystal $a$-, $b$-, $c$-axis \cite{fu}, so the 90$^\circ$ peak corresponds to the NMR component with $H_0$ // $ab$ plane.

\begin{figure}[htbp]
\includegraphics[width=6cm]{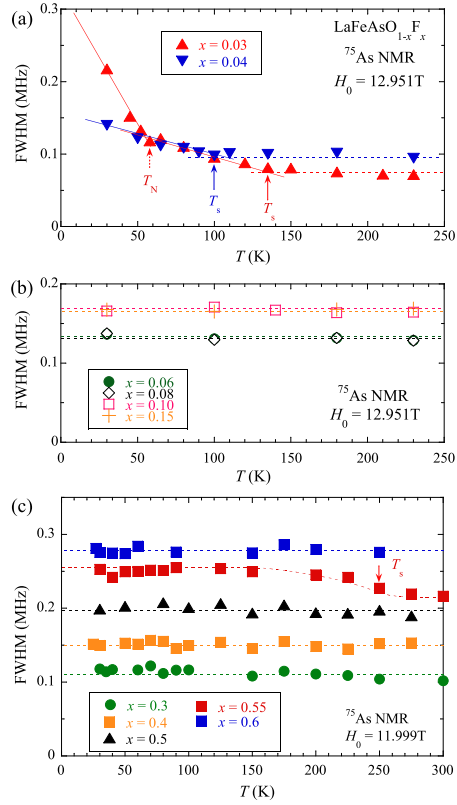}
\caption{(color online) The temperature dependence of the FWHM for 90$^\circ$ peak of $^{75}$As NMR spectrum.
Solid and dotted arrows indicate $T_{\rm s}$ and $T_{\rm N}$, respectively.
}
\label{AsNMR_FWHM}
\end{figure}

\begin{figure}[htbp]
\includegraphics[width=6cm]{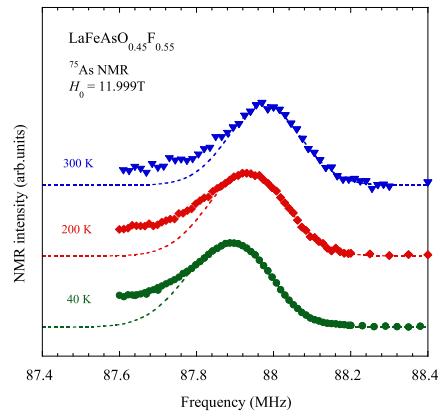}
\caption{
The temperature dependence of $\theta$ = 90$^{\circ}$ peak of $^{75}$As NMR spectrum for LaFeAsO$_{0.45}$F$_{0.55}$. The dotted lines are Gaussian function fittings.
}
\label{AsNMR_HP}
\end{figure}

Figure \ref{AsNMRspectrum} (b) and (c) enlarge the $\theta$ = 90$^\circ$ peak for $x$ = 0.03 and 0.04, respectively.
The spectra of both $x$ = 0.03 and 0.04 are broadened gradually as the temperature is lowered. To see this in more datail, we plot the temperature dependence of FWHM of the 90$^\circ$ peak for $x$ = 0.03 and 0.04 in Fig. \ref{AsNMR_FWHM}(a), which are obtained by Gaussian fittings to the spectra.
The FWHM of $x$ = 0.03 shows an anomaly at 135 K, followed by a steeper increase below 58 K where an AF order sets in.
For $x$ = 0.04, the FWHM keeps constant at high temperatures but increases below 100 K. By contrast, the FWHM is temperature independent from 30 K to 230 K for 0.06 $\leq$ $x$ $\leq$ 0.15, as shown in Fig. \ref{AsNMR_FWHM}(b).

Below we illustrate that $T$ = 135 K for $x$ = 0.03
and $T$ = 100 K for $x$ = 0.04 correspond to a structural phase transition temperature $T_{\rm s}$.
For single crystal LaFeAsO, $^{75}$As NMR spectrum with $H_0$ // ab plane shows a single peak above $T_{\rm s}$ and this peak splits into two corresponding to  $H_0$ // $a$ axis and $H_0$ // $b$ axis below $T_{\rm s}$\cite{fu}.
This is because orthorhombic distortion breaks the fourfold ($C4$) rotation symmetry of the EFG and the second order effect of the nuclear quadrupolar interaction in $H_0$ // $a$ axis differ from that in  $H_0$ // $b$ axis.
Since the difference of the second order effects was smaller, in pollycrystalline samples we observed only the increase of FWHM below $T_{\rm s}$, rather than the split of the spectrum.

%Figure 6 (a) and (b) show the $^{139}$La-NQR spectra of $x$=0.03 measured at $T$=11K and 0.06 measured at $T$=30K, respectively.
%Although only three peaks which correspond to $I$=9/2 were observed in $x$=0.06, six peaks were observed in $x$=0.03.
%These consistent with the result that there are two site (Low, High) in $x$=0.03 revealed by  $^{75}$As-NQR.
%%$\nu_{Q}$ and  asymmetry parameter $\eta$ can be calculated from the peak positions.
%For $x$=0.06, $\nu_{Q}$ and  $\eta$ are calculated to be 1.25MHz and 0.12.
%For $x$=0.03, $\nu_{Q}$ and  $\eta$ of one site are calculated to be 1.35MHz and 0.18,  $\nu_{Q}$ and  $\eta$ of the other site are calculated to be 1.58MHz and 0.25.
%In contradiction to $^{75}$As, $\nu_{Q}$ of $^{139}$La decreases with increasing $x$\cite{nakai1}.
%Each of $\eta$ for $x$=0.03 is larger than that of $x$=0.06.
%This is consistent with the occurrence of structural phase transition occurs in $x$=0.03.
%% is because structural phase transition occurs at $T$=11K in $x$=0.03.

%\begin{figure}[h]
%\includegraphics[width=8.0cm]{LaNQRspectrum2.eps}% Here is how to import EPS art
%\caption{\label{Fig6} (color online) $^{139}$La-NQR spectrum above $T_{\rm c}$ for $x$=0.03 (a) and 0.06 (b).
%}
%\end{figure}

For $x > 0.5$,  we have found a structural phase transition  by measuring the asymmetry of the EFG in the previous work\cite{yang}. Here we show that, for HP samples, the FWHM  of NMR spectra recognized the structural phase transition as well. Figure \ref{AsNMR_HP} shows the $\theta$ = 90$^\circ$ peak at some typical temperatures for $x$ = 0.55. The spectra become broad with decreasing temperature.
Figure \ref{AsNMR_FWHM}(c) exhibits the temperature dependence of the FWHM of 90$^\circ$ peak for HP samples obtained by Gaussian fittings to the spectra.
For $x$ = 0.3-0.5, FWHM are temperature independent. For $x$ = 0.55, FWHM shows an abrupt change at around $T$ = 250 K, which corresponds to the structural phase transition temperature $T_{\rm s}$ as the case of $x$ = 0.03 and 0.04.
%, and the latter corresponds to $T_{\rm c}$, below which the spectra are broadened due to the field distribution in the vortex state. For $x$ = 0.6, FWHM also increases below $T_{\rm c}$.
It is worth noting that we find no sign of AF order from NMR spectra for HP LaFeAsO$_{1-x}$F$_{x}$ samples. %, as will be shown in detail in the next section.

\begin{figure}[htbp]
\includegraphics[width=8.0cm]{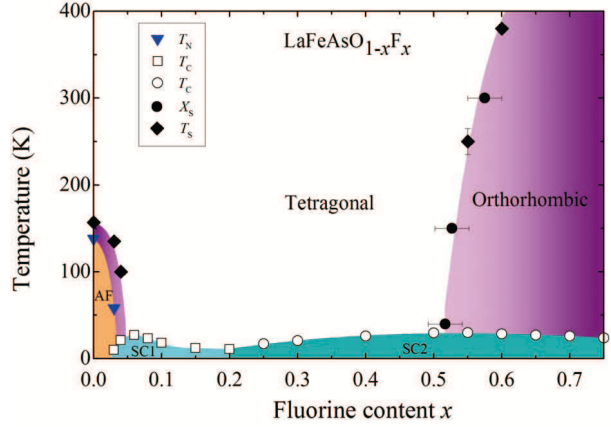}
\caption{(color online) The complete phase diagram for LaFeAsO$_{1-x}$F$_{x}$. AF denotes the antiferromagnetically ordered phase, SC1 and SC2 denote the superconducting domes obtained by conventional solid-state and high-pressure  synthesis methods, respectively.
$T_{\rm s}$ and $T_{\rm N}$ for $x$ = 0 are referred from ref.\cite{Huang}. The values of $x_s$ are from NMR measurements\cite{yang}.
}
\label{phasediagram_a}
\end{figure}

The $T_{\rm c}$, $T_{\rm s}$ and $T_{\rm N}$ for the entire $x$ range  are summarized in the phase diagram shown in Fig. \ref{phasediagram_a}.
The $T_{\rm c}$ forms two superconducting domes peaked at $x_{opt}$ = 0.06 with $T_{\rm c}$ = 27 K and  $x_{opt}$=0.5$\sim$0.55 with $T_{\rm c}$ = 30 K, respectively.
Above $T_{\rm c}$, the TEM images together with the NMR spectra, evidence that a $C4$ symmetry-breaking structural phase transition takes place above both domes, with $T_{\rm s}$ varying strongly with $x$.
In the first dome, the suppression of $T_{\rm s}$ and $T_{\rm N}$ shows a second-order-like variation towards superconducting dome, while $T_{\rm s}$ intersects the second dome. This is the first report showing that the phase diagram of LaFeAsO$_{1-x}$F$_{x}$ at low doping region is actually similar to that of 122 FeSCs.

\subsection{Possible new type of quantum criticality in the second dome}\label{QCP}%%%%%%%%%%%%%%%%%%%%%%%%%%%%%%%%

\begin{figure}[htbp]
\includegraphics[width=9cm]{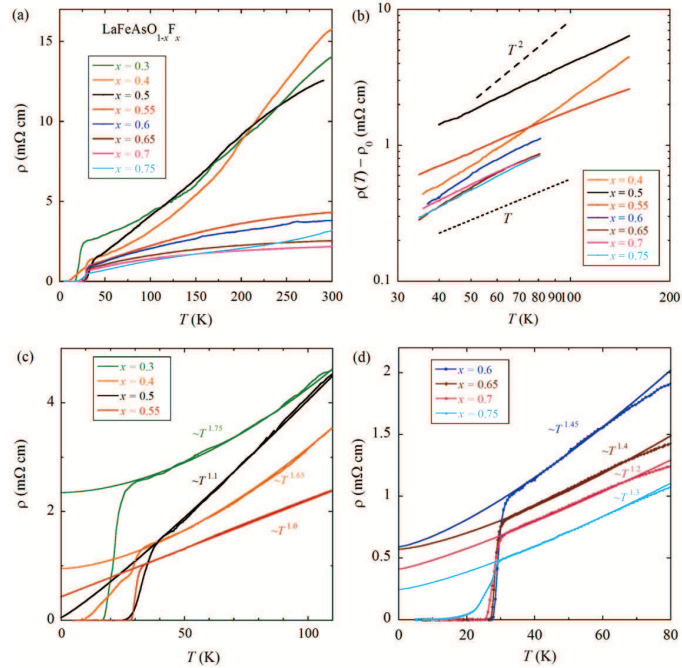}
\caption{(a)The electrical resistivity for HP LaFeAsO$_{1-x}$F$_{x}$ with 0.3 $\leq x \leq$ 0.75. (b)Log[$\rho(T) - \rho_0$] vs. log $T$ plots. The dashed and dotted lines are guides to the eyes showing $\rho(T)$ $\sim$ $T^2$ and $T$, respectively. (c)(d)The low temperature electrical resistivity. The solid lines are the fittings to $\rho = {{\rho }_{0}}+A{{T}^{n}}$ over the temperature range shown.
}
\label{RT}
\end{figure}

In this section, we discuss possible new type of quantum criticality in the high-doped region on the basis of resistivity measurement. The previous works reported that the parent compound LaFeAsO and the underdoped samples show a kink in electrical resistivity $\rho$ due to structural or magnetic transitions\cite{cruz,Dong}. In addition, $\rho$ obeys $\rho(T) \sim T^2$ variation in the first dome for $x$ $\leq$ 0.2 \cite{nakai_rho,Hess}. %, as described by Fermi liquid theory
The temperature dependence of $\rho$ for the second dome (0.3 $\leq x \leq$ 0.75) are presented in Fig.\ref{RT}(a). For all samples, no anomaly is observed over the temperature range from $T_{\rm c}$ to 300 K, which implies that AF order is absent in the second dome, being consistent with the NMR/NQR spectra. We fitted the normal state resistivity to $\rho = {{\rho }_{0}}+A{{T}^{n}}$ and obtained the residual resistivity $\rho_0$, the coefficient $A$ and the exponent $n$.
Figure \ref{RT}(c)(d) show the fittings over the temperature range shown.

\begin{figure}[tbp]
\includegraphics[width=8.0cm]{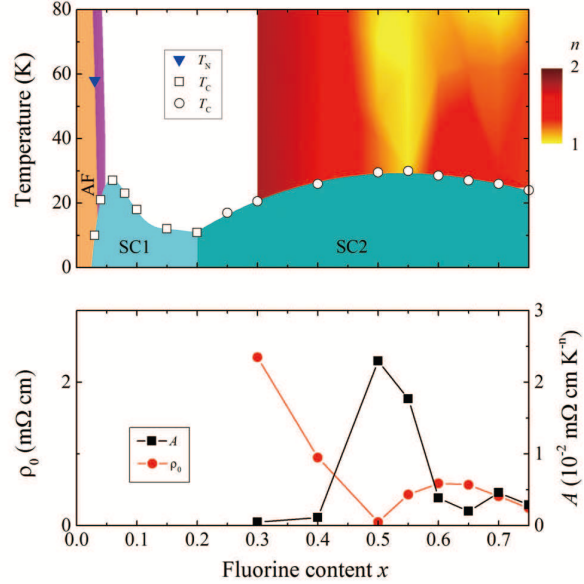}
\caption{(color online)
(a) The phase diagram for LaFeAsO$_{1-x}$F$_{x}$ obtained by present work. The evolution of the exponent $n$  is obtained by $\rho = {{\rho }_{0}}+A{{T}^{n}}$.
(b) The doping dependence of $\rho_0$ and $A$.
}
\label{phasediagram_b}
\end{figure}

\begin{figure}[htbp]
\includegraphics[width=7.5cm]{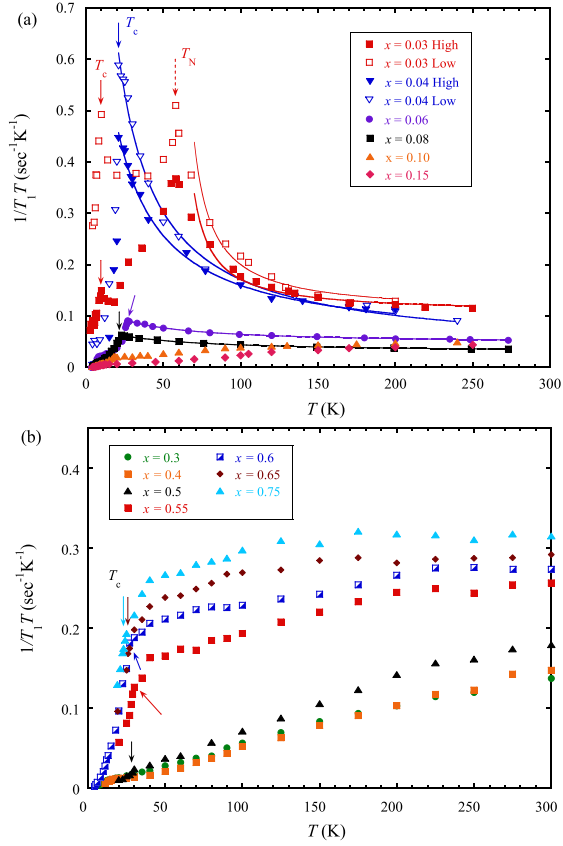}
\caption{ (a)(b)The $^{75}$As nuclear spin-lattice relaxation rate divided by temperature $1/T_{1}T$ for AP and HP LaFeAsO$_{1-x}$F$_{x}$, respectively. The solid lines are fittings to $1/T_{1}T=(1/T_{1}T)_{0}+C/(T+\theta)$\cite{moriya1}, where $(1/T_{1}T)_{0}$ is the contribution from the density of states at the Fermi level and $C/(T+\theta)$ describes the contribution from  antiferromagnetic fluctuations. The dashed arrow indicates $T_{\rm N}$ for $x$ = 0.03. The solid arrows indicates $T_{\rm c}$ for corresponding $x$ concentrations.
}
\label{T1T}
\end{figure}

In contrast to the first dome, $\rho$ of the second dome shows a non-Fermi liquid (NFL) behavior with $n<2$. In particular, $\rho$ shows a $T$-linear behavior with $n$ = 1 at $x$ = 0.55. To compare the variation of $n$ more intuitively, we plot $\rho(T) - \rho_0$ versus $T$ in logarithmic coordinates, as shown in Fig.\ref{RT}(b).
The evolution of $n$, $\rho_0$ and $A$ with F content are presented in Fig.\ref{phasediagram_b}.
The $\rho_0$ value of HP samples is comparable to that of  AP samples\cite{Dong}. %, but is one order of magnitude larger than that of 122 single crystal samples\cite{Zhou}.
Since $\rho_0$ is a measure of the disorder degree, the results indicate that the quality of HP samples is close to that of the AP samples. %, and it is reasonable that the single crystal sample has a better crystallization.
%Furthermore, $\rho_0$  shows a trend of decrease with increasing $x$.
%%, which is an indication of the system being away from a magnetic phase.
%The larger $\rho_0$ in the samples close a magnetic phase can be understood as due to impurity scattering enhanced by spin fluctuations \cite{Miyake,Kontani}, and $\rho_0$ decreases when the spin fluctuations become weak.

At the optimal doping level $x$ = 0.5$\sim$0.55 where $T_{\rm c}$ is maximal and  $n$ = 1 is observed,
the coefficient $A$ also shows maximum. $A$ is proportional to  $(m^{\ast})^{2}$, where $m^{\ast}$ is the effective electron mass. These features are often considered as the signature of a magnetic QCP\cite{moriya}. However, the second dome is far away from an AF order and no low-energy spin fluctuations can be found. Figure \ref{T1T} shows the $^{75}$As $1/T_{1}T$ for AP and HP LaFeAsO$_{1-x}$F$_{x}$. In the low doping regime close to the AF ordered phase, $1/T_{1}T$ increases rapidly with decreasing temperature (Fig. \ref{T1T} (a)), while such increase is absent in the second dome (Fig. \ref{T1T} (b)).
The results indicate the presence of strong AF spin fluctuations  in the first dome, whereas neither AF order nor low-energy spin fluctuation can be found in the second dome. Considering the fact that $T_{\rm s}$ extrapolates to zero at $x_{opt}$, it is more likely that the  $T$-linear behavior of the resistivity arises from quantum criticality associated with the structural phase transition. It has been found that below $T_{\rm s}$ electronic nematicity appears\cite{JHChu,Lu,Zhou-PRB,Kasahara,Yi}. Theoretically, it was also shown that electronic nematic QCP can lead to NFL behavior\cite{Kivelson}.

%In the first dome, the suppression of $T_{\rm s}$ and $T_{\rm N}$ shows a second-order-like variation towards superconducting dome, while  antiferromagnetism and superconductivity coexist  microscopically for $x$ = 0.03. These features are not seen in the previous reports\cite{Luetkens,Huang} and imply  a magnetic QCP may exist in proximity to the AF phase. Indeed, the evolution of the AF spin excitations with doping level $x$ measured by nuclear spin-lattice relaxation rate divided by $T$ ($1/T_{1}T$) gives evidence\cite{Oka}, as shown in Fig.\ref{T1T}(a).

In fact, a two-superconducting-dome phase diagram has also  been  found in  LaFeAsO$_{1-x}$H$_{x}$\cite{Hiraishi_LaFeAsOH}, K$_{1-x}$Fe$_{2-y}$Se$_{2}$\cite{Sun} and  LaFeAs$_{1-x}$P$_{x}$O\cite{LaFeAsPO},  and  more recently, also in K-doped FeSe thin films\cite{Xue}. However, our system is quite different from others.
The two domes in LaFeAsO$_{1-x}$H$_{x}$, LaFeAs$_{1-x}$P$_{x}$O and K$_{1-x}$Fe$_{2-y}$Se$_{2}$ are all closely adjacent to a magnetic ordered state\cite{LaFeAsPO,Hiraishi_LaFeAsOH,Guojing}, while in the higher-$T_{\rm c}$ superconducting dome of K-doped FeSe thin film neither magnetism nor structural phase transition is present\cite{Xue}.
%The two domes  differ in magnetic and  superconducting properties, and more importantly, electronic structures of the two domes may be different considering the multi-orbital physics in FeSCs.
%It is worthy to be reminded that the phase diagram of CeCu$_{2}$Si$_{2}$ also has two superconducting phases, one close to magnetism and the other far away from magnetism\cite{Yuan}.
%The nature of QCP and associated quantum critical fluctuation is one of the key issues to understand the origin of two domes.
Thus the present system offers a unique opportunity to study the quantum
criticality due to AF order and possibly electronic nematic order  simultaneously.

\section{conclusion}%%%%%%%%%%%%%%%%%%%%%%%%%%%%%%%%%%%%%%%%%%%%%%%%%%%%%%%%%%%%%%%%%%%%%%%%%%%%%%%%%

In conclusion, we have performed   measurements on LaFeAsO$_{1-x}$F$_x$ ($x$ = 0.03-0.75)
by NMR  and TEM. We demonstrated that a similar $C4$-symmetry-breaking structural phase transition takes place in the two doping regimes where two superconducting domes are formed.
In the low-doping regime of $x\leq$0.2, $T_{\rm s}$ and $T_{\rm N}$ are well separated, and both show a second-order-like suppression with increasing doping level.
For the  $x$ = 0.03 sample, we find that $^{75}$As nuclear spin-lattice relaxation rate 1/$T_1$ shows a clear peak at $T_{\rm N}$ = 58 K due to a critical slowing down of the magnetic moment and then a further decrease below $T_{\rm c}$ = 9.5 K, which indicates that AF order and superconductivity coexist microscopically. % in LaFeAsO$_{0.97}$F$_{0.03}$.
Furthermore, 1/$T_1$ below 0.6$T_{\rm c}$ %for $x$ = 0.03
decreases in proportion to  $T$, indicating gapless excitations in the coexisting state.
%Such temperature dependence of $1/T_1$ in the coexistence region is similar to the cases of  Ba$_{1-x}$K$_{x}$Fe$_2$As$_2$ and Ca$_{1-x}$La$_{x}$FeAs$_{2}$.
In the second dome, by contrast, there is neither  AF order nor low-energy spin fluctuations. The $T_{\rm s}$ extrapolates to zero at around $x_{opt}$ = 0.5$\sim$0.55, where $T_{\rm c}$ is the maximal. The $T$-linear behavior of electrical resistivity  and the maximum of coefficient $A$ seen at $x_{opt}$ points to  a new type of quantum criticality which may provide a new route to high temperature superconductivity.

\section*{ACKNOWLEDGMENTS}
We thank Z. X. Zhao for useful communication and S. Kawasaki for help in some of the measurements. This work was supported by the National Key R$\&$D Program of China (No. 2017YFA0302904), the National Natural Science Foundation of China (No. 11674377, No. 11634015), and the Chinese Academy of Sciences (No. XDB07020200). Work in Okayama was supported by research grants from JSPS (No. 16H0401618).

%\bibliography{apssamp}

\end{document}